\documentclass[aps,prx,bibliography,superscriptaddress,,twocolumn]{revtex4-2}
\usepackage[utf8]{inputenc}
\usepackage{amsmath}
\usepackage{physics}
\usepackage{mathtools}
\usepackage{amsfonts}
\usepackage{amssymb}
\usepackage{braket}
\usepackage{graphicx}
\usepackage{subfigure}
\usepackage{textcomp}
\usepackage{color}
\usepackage{float}
\usepackage[dvipsnames]{xcolor}
\usepackage{mathrsfs}
\usepackage{natbib}
\usepackage{makecell}
\usepackage{soul}

\usepackage{multirow}

\begin{document}
\title{Field-Tunable Quantum Metric in Few-Layer Phosphorene}
\author{Md Afsar Reja}
\email{afsarmd@iisc.ac.in}
\affiliation{Solid State and Structural Chemistry Unit, Indian Institute of Science, Bangalore 560012, India}
\author{Arka Bandyopadhyay}
\email{arka.bandyopadhyay@uni-wuerzburg.de}
\affiliation{Institute for Theoretical Physics and Astrophysics, University of Würzburg, 97074 Würzburg, Germany}
\author{Awadhesh Narayan}
\email{awadhesh@iisc.ac.in}
\affiliation{Solid State and Structural Chemistry Unit, Indian Institute of Science, Bangalore 560012, India}
\date{\today}

\begin{abstract}
The quantum metric -- which quantifies the distance between quantum states -- is a fundamental component of the quantum geometric tensor, playing a crucial role in a wide range of physical phenomena. Its direct detection and control remains a challenge, requiring suitable material candidates. In this work, we present the emergence of a tunable quantum metric in a versatile two-dimensional material platform, namely, few-layer phosphorene. Using ab-initio-derived models, we show how electric fields can be used to substantially enhance the quantum metric as well as the associated quantum weight. Furthermore, we present a layer-dependent evolution of the quantum metric and its interplay with the electric field in this material. Our results establish few-layer phosphorene as a promising platform for exploring control over the quantum metric and the resulting metric responses in real materials.
\end{abstract}

\maketitle

\noindent\textit{\textbf{Introduction--}} 
Geometry of the Hilbert space -- termed quantum geometry -- plays an important role in understanding various phenomena in quantum matter~\cite{provost1980riemannian,cheng2010quantum,yu2024quantum,liu2025quantum,torma2023essay}. The imaginary part of the quantum geometric tensor (QGT), known as the Berry curvature, has been widely studied, especially for its connection to the quantum Hall effect~\cite{xiao2010berry} and non-linear transport effects~\cite{bandyopadhyay2024non}. The real part of the QGT is the quantum metric tensor (QMT), which geometrically represents the distance between two quantum states~\cite{provost1980riemannian}. In recent years, the quantum metric has garnered considerable attention for its impact on a wide range of physical phenomena, including flat-band superconductivity~\cite{peotta2015superfluidity,xie2020topology,herzog2022superfluid}, localization, and transport effects~\cite{gao2014field,liu2021intrinsic,wang2021intrinsic,gao2023quantum,liu2025quantum,yu2024quantum,das2023intrinsic}.

Phosphorus, one of earth’s most abundant elements, exists in several distinct allotropes~\cite{deringer2020hierarchically}. Among these allotropes, black phosphorus stands out as the thermodynamically stable form. It possesses a distinctive layered orthorhombic structure, enabling mechanical or chemical exfoliation down to the monolayer limit -- commonly known as phosphorene~\cite{liu2014phosphorene,li2014black,woomer2015phosphorene,wu2021large,ding2024synthesis}. Unlike other two-dimensional materials, it uniquely combines a tunable, layer-dependent direct band gap with high carrier mobility and pronounced in-plane anisotropy~\cite{liu2014phosphorene,tran2014layer,liu2015semiconducting,bandyopadhyay2023berry}. Few-layer phosphorene has further emerged as a versatile quantum material, capable of undergoing a transition from a moderate-gap semiconductor to a band-inverted semimetal under external perturbations such as pressure, electric fields, and strain~\cite{kim2015observation,xiang2015pressure,liu2015switching,dolui2015quantum}. These remarkable tunable properties position phosphorene as a promising avenue for exploring band geometric phenomena.

\begin{figure}[b]
\centering
\includegraphics[width=0.48\textwidth]{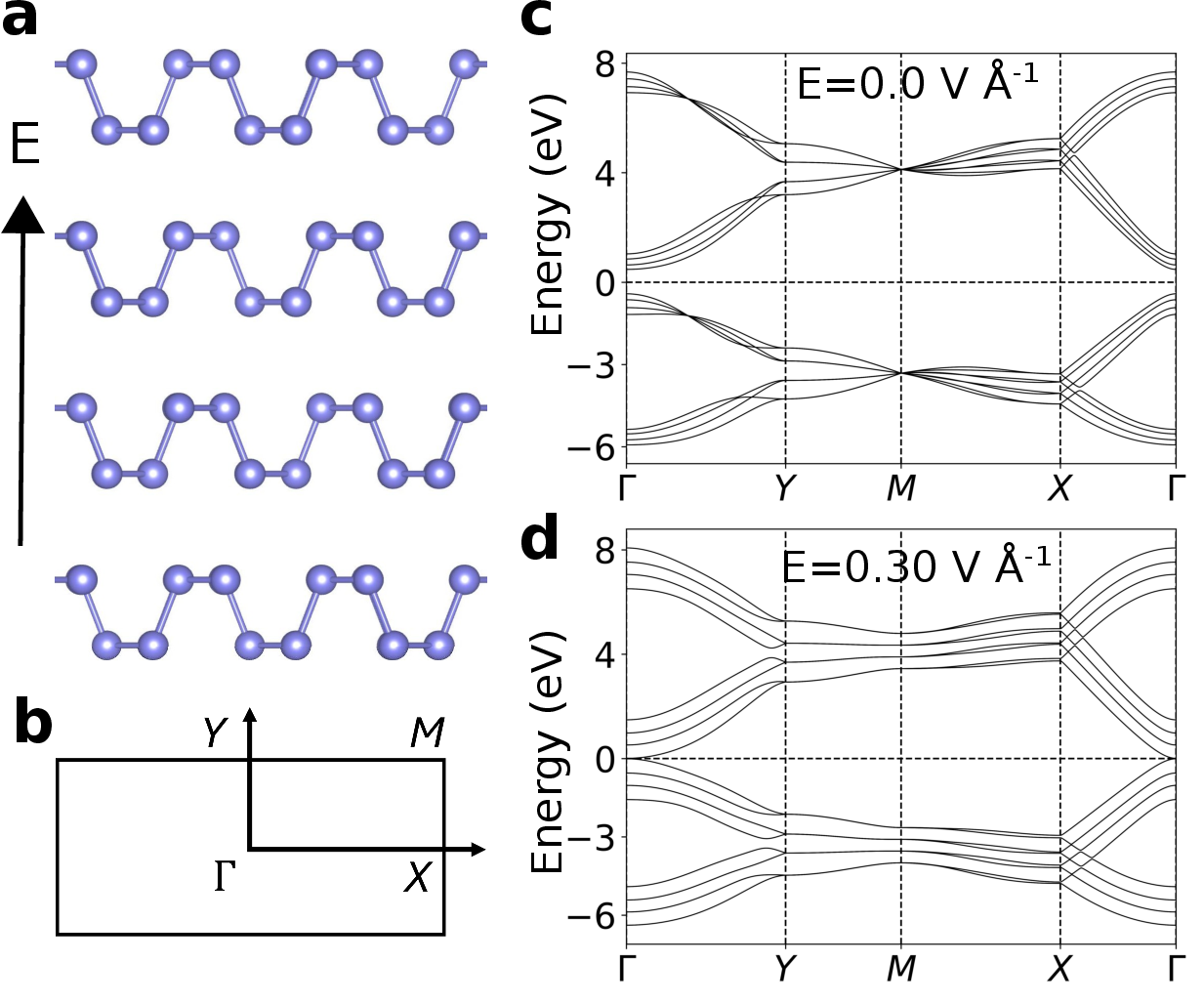}
\caption{\textbf{Crystal structure and electronic band structure of few-layer phosphorene.} (a) Side view of the few-layer phosphorene showing four layers. The electric field, $E$, shown by the black arrow, is applied perpendicular to the layers along the stacking direction. (b) The Brillouin zone with the high symmetry points shown. Band structure of four-layer phosphorene (c) without and (d) with an electric field $E = 0.30~\mathrm{V\,\AA^{-1}}$. We note that with increasing electric field, the band gap decreases and closes at a critical value.}
\label{Fig_4layer_bands_E}
\end{figure}

The QMT has been experimentally explored in a handful of platforms, including superconducting qubits~\cite{tan2019experimental}, nitrogen vacancy centers~\cite{yu2020experimental}, and in very few solid-state materials~\cite{kang2025measurements,sala2025quantum}. However, most experiments so far have inferred the quantum metric components indirectly, for example, through measurements of nonlinear transport phenomena~\cite{wang2023quantum,gao2023quantum,yu2025quantum,liu2025giant}. Recently, however, the focus has shifted toward direct measurements of the QMT. A quasi-quantum metric has been measured in the kagome lattice material CoSn~\cite{kang2025measurements}. In a recent study, Kim \textit{et al.} have extracted the quantum metric components in black phosphorus by modeling it as an effective two-level system and measuring the corresponding pseudospins~\cite{kim2025direct}.

Motivated by these recent developments, in this work, we propose few-layer phosphorene as a versatile material platform for exploring the QMT. We discover that both a layer-dependent band gap tunability and an electric-field-induced band gap modulation play crucial roles in determining the magnitude and distribution of the quantum metric components. Specifically, for quadruple-layer phosphorene, we show that the magnitude of QMT components is enhanced significantly with applied electric fields. We also demonstrate that the tunability of the quantum metric is possible in bi-layer and tri-layer cases. Furthermore, we show that the applied electric field plays a crucial role in the behavior of the quantum weight. Overall, our work introduces a promising platform for exploring quantum metric in real materials. \\

\begin{figure}[t]
% \centering
\includegraphics[width=0.48\textwidth]{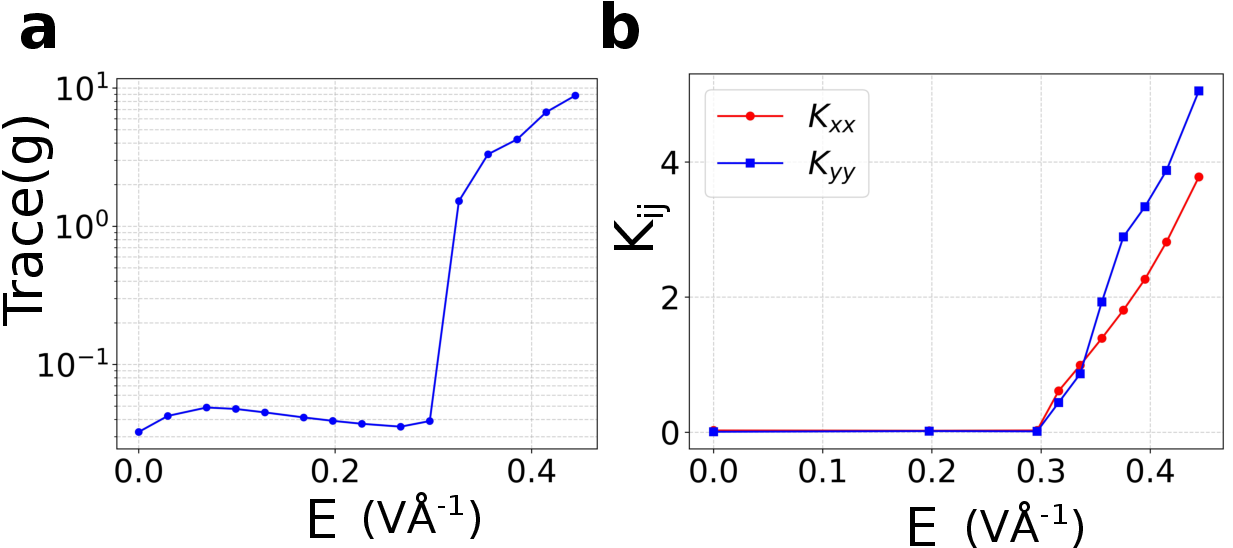}
\caption{\textbf{Tunable quantum metric in quadruple-layer phosphorene with varying electric field.} (a) Trace of the quantum metric for quadruple-layer phosphorene with increasing electric field, $E$. (b) Variation of the quantum weight components $K_{xx}$ and $K_{yy}$ with the electric field. Note the pronounced increase near the critical electric field, $E\approx 0.3~\mathrm{V\,\AA^{-1}}$. The quantum weight and the quantum metric components can be directly tuned by an applied electric field.}
\label{Fig_QW_Trace_tetra}
\end{figure}

\noindent\textit{\textbf{Model and Computational Methods--}}
Similar to graphite, black phosphorus [Fig.~\ref{Fig_4layer_bands_E}(a)] is a layered material, and, in analogy with graphene, its monolayer form -- phosphorene -- has attracted significant interest due to its distinctive structural and electronic properties. In phosphorene, the atoms are arranged in a two-dimensional hexagonal lattice that exhibits a puckered configuration along the armchair direction (see Appendix for details). The rectangular unit cell of monolayer phosphorene contains four atoms, and the two lattice vectors are given as $a = 4.38~\text{\AA}$ and $b = 3.31~\text{\AA}$. The corresponding first Brillouin zone, shown in Fig.~\ref{Fig_4layer_bands_E}(b), highlights the high-symmetry points. We formulate the tight-binding Hamiltonian for the layered phosphorene material, illustrated in Fig.~\ref{Fig_4layer_bands_E}(a), as

\begin{equation}
\mathcal{H} = \sum_{i} \varepsilon_{i} c^{\dagger}_{i} c_{i} 
  + \sum_{i \neq j} t_{ij} c^{\dagger}_{i} c_{j} 
  + \sum_{\alpha} V_{\alpha} c^{\dagger}_{\alpha} c_{\alpha},
  \label{eq:TB-Hamiltonian}
\end{equation}

where $\varepsilon_{i}$ denotes the onsite energy at site $i$ in the absence of any external perturbation. It is assumed to be uniform, i.e., $\varepsilon_i = \varepsilon, \; \forall ~i$, which further determines the Fermi level of the system. The term $t_{ij}$ denotes the hopping amplitude between the $i$-th and $j$-th atomic sites ($i \neq j$), which can be further categorized into intralayer ($t^{\parallel}$) and interlayer ( $t^{\perp}$) hopping components. The application of a transverse electric field, depicted by the black arrow in Fig.~\ref{Fig_4layer_bands_E}(a), gives rise to additional potential energies for atoms located at different heights along the field direction. These potentials, denoted by $V_{\alpha}$, vary across the atomic layers indexed by $\alpha$ along the $z$ direction. The complete set of hopping parameters for layered phosphorene, considered in this work, is listed in Table~\ref{tab:hopping-parameters} (see Appendix). To benchmark the tight-binding model parameters of the Hamiltonian given in Eq.~\ref{eq:TB-Hamiltonian}, we compared the band structure of monolayer phosphorene with the \textit{ab-initio} dispersion obtained using the accurate Heyd-Scuseria-Ernzerhof hybrid functional, 
which predicts a direct band gap of $\approx 1.5$~eV at the $\Gamma$ point~\cite{qiao2014high,bandyopadhyay2023berry,rudenko2014quasiparticle} -- our tight-binding model closely matches this value.

The QGT is a complex-valued tensor defined as~\cite{provost1980riemannian,ma2010abelian}, 

\begin{equation}
Q^{n}_{ij}(\mathbf{k}) = \braket{ \partial_{k_i} \psi_n(\mathbf{k}) | \left(1 - P_n(\mathbf{k})\right) | \partial_{k_j} \psi_n(\mathbf{k}) }, 
\label{Eq_QGT}
\end{equation}

where, $i$, $j$ represent the spatial indices, \( \partial_{k_i} = \frac{\partial}{\partial k_i} \), $n$ is the band index, and the projection operator is given by \( P_n(\mathbf{k}) = \ket{\psi_n(\mathbf{k})} \bra{\psi_n(\mathbf{k})} \). The real part of the QGT corresponds to the QMT, defined as $g^{n}_{ij}(\mathbf{k}) = \mathrm{Re}\!\left[ Q^{n}_{ij}(\mathbf{k}) \right]$, while its imaginary part gives the Berry curvature, $\Omega^{n}_{ij}(\mathbf{k}) = -2\,\mathrm{Im}\!\left[ Q^{n}_{ij}(\mathbf{k}) \right]$. Another related quantity is the quantum weight $K$, introduced in Ref.~\cite{onishi2024quantum}, which plays a central role in determining the dielectric response and optical properties of materials. Its components are given by the integral over the Brillouin zone of the quantum metric components,

\begin{equation*}
    K_{ij} = 2\pi \int \frac{d^2 k}{(2\pi)^2} \, g_{ij}(\mathbf{k}).
\end{equation*}

In particular, we compute the quantum weights $K_{xx}$ and $K_{yy}$ associated with the corresponding components of the quantum metric for few-layer phosphorene. \\

\begin{figure*}[t]
\centering
\includegraphics[width=0.90\textwidth]{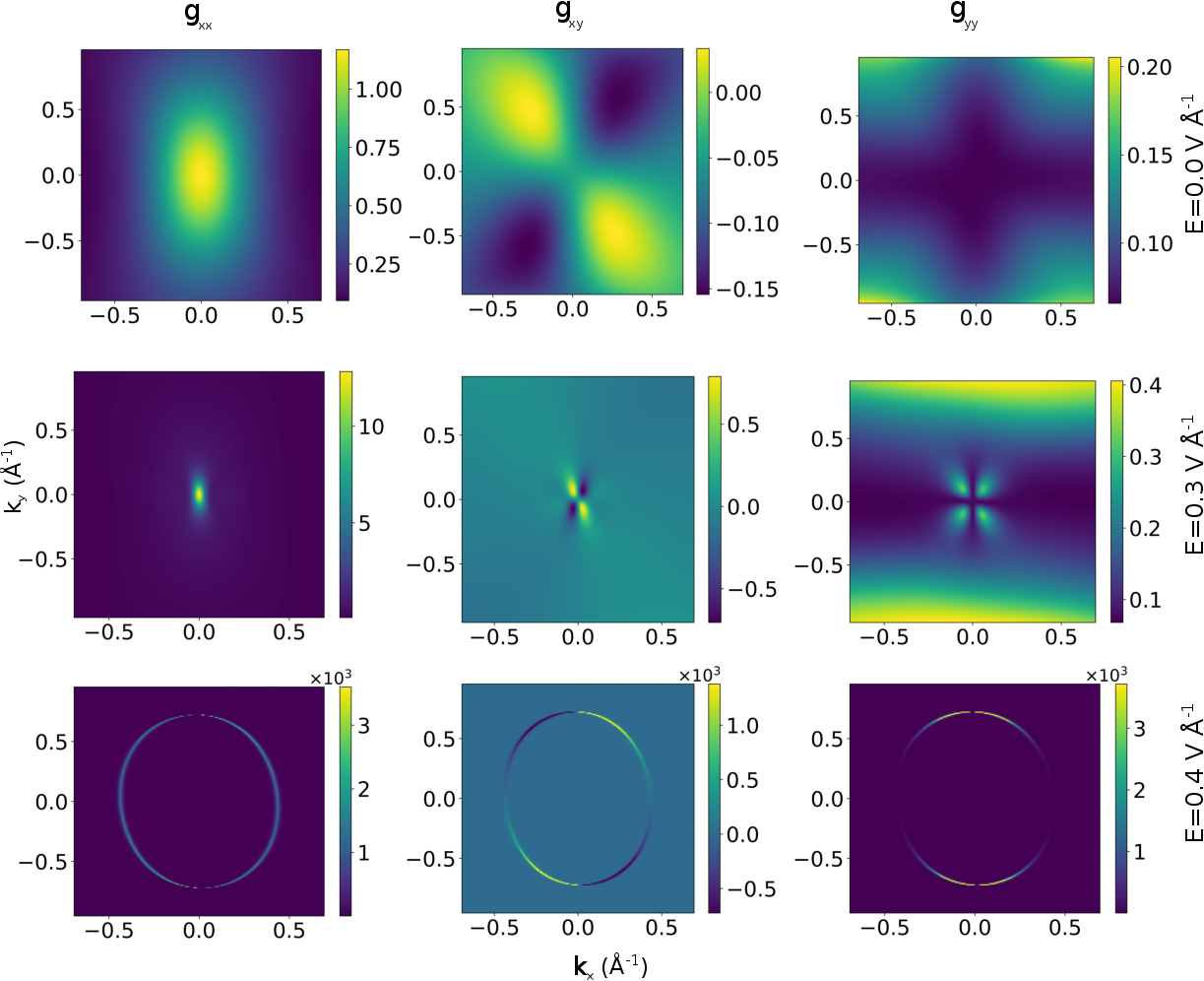}
\caption{\textbf{Distribution of the quantum metric components of quadruple-layer phosphorene under varying electric field.} The quantum metric components $g_{xx}$, $g_{xy}$, and $g_{yy}$ are plotted column-wise, while the electric field increases vertically downward. We note that with increasing electric field, each component attains large values with order of magnitude enhancements seen near $E = 0.30~\mathrm{V\,\AA^{-1}}$.}
\label{Fig_4layer_QMT}
\end{figure*}

\noindent\textit{\textbf{Tunable quantum metric in quadruple-layer phosphorene--}} The electronic band structure of monolayer phosphorene, obtained from the tight-binding Hamiltonian presented in Eq.~\ref{eq:TB-Hamiltonian}, reveals a distinct direct band gap of 1.52 eV at the $\Gamma$ point, characteristic of its semiconducting behavior. As the number of layers increases, additional interlayer couplings lead to a progressive narrowing of this gap. Specifically, the band gap exhibits a substantial reduction to $1.12$ eV, $0.97$ eV, and $0.89$ eV for the bi-, tri-, and quadruple-layer phosphorene systems, respectively. Remarkably, the application of a transverse electric field induces a pronounced modulation of the electronic band structure, particularly in the vicinity of the Fermi level, across all these layered configurations. With increasing field strength, the band gap systematically narrows down and ultimately vanishes at a critical electric field, marking an electric-field-driven semiconductor-to-metal transition~\cite{kim2015observation,liu2015switching,dolui2015quantum,bandyopadhyay2023berry}. Motivated by this general behavior of these layered systems -- namely, the electric-field–induced modulation of the band gap -- we focus on the quadruple-layer phosphorene as a representative case and discuss its properties in detail, noting that the other layered systems exhibit a similar trend.

We begin with the band structure of quadruple-layer phosphorene, presented in Fig.~\ref{Fig_4layer_bands_E}(c) and Fig.~\ref{Fig_4layer_bands_E}(d), without and with an applied electric field ($E = 0.30~\mathrm{V\,\AA^{-1}}$), respectively. We clearly see the effective tunability of the band gap, which can even be reduced to zero upon applying an external electric field. We have utilized this band gap tunability in few-layer phosphorene to control the QMT. First, we calculate the QMT components for each band using a gauge invariant formulation employing the Kubo formulae (see Appendix for details). We evaluate the trace of the QMT for the valence band by integrating $(g_{xx} + g_{yy})$ over the Brillouin zone for different applied electric fields, $E$. We investigate the trends of the quantum metric components for the valence band, as this band plays a crucial role in determining transport properties. We plot the trace of QMT for the valence band in Fig.~\ref{Fig_QW_Trace_tetra} (a). We find that the trace stays far below one until the critical electric field. It exhibits an abrupt jump to unity at $E \approx 0.3~\mathrm{V\,\AA^{-1}}$, and thereafter increases by nearly an order of magnitude with further variation in $E$. The variations of $K_{xx}$ and $K_{yy}$ with the electric field are shown in Fig.~\ref{Fig_QW_Trace_tetra}(b). We observe that the quantum weights follow the same trend as the trace of the QMT, as expected. Below the critical electric field, both $K_{xx}$ and $K_{yy}$ are near zero, and beyond the critical value of $E$, they continue to increase steadily with increasing $E$, indicating an enhancement of the underlying quantum metric as the band gap is reduced. This reveals a direct tunability of the QMT in our proposed system using an applied electric field.

To further understand the above features of QMT under an applied $E$, we next examine the distribution of the QMT components in the Brillouin zone. We plot the individual QMT components for the valence band in Fig.~\ref{Fig_4layer_QMT} for different applied electric fields. The distribution of $g_{xx}$, $g_{xy}$, and $g_{yy}$ at $E = 0.0$ closely matches the measured and calculated results obtained in Ref.~\cite{kim2025direct} for bulk black phosphorus. However, upon switching on the electric field, the distribution as well as the magnitudes of the QMT components change drastically. For instance, with increasing $E$ the band gap reduces near $\Gamma$. Concomitantly, we find that the $g_{xx}$ values get enhanced by an order of magnitude, and the distribution also shrinks toward the $\Gamma$ point, as a result of the sharpening of the valence band dispersion. Beyond the critical $E$, a band inversion occurs resulting in the valence and conduction bands crossing along a ring-like feature. In this case, we obtain substantially enhanced values of $g_{xx}$ along this ring where the band gap closes. We note that upon including spin-orbit coupling a small gap will open along this ring~\cite{liu2015switching,dolui2015quantum}. Beyond the critical field, our calculations reveal that the magnitudes of the QMT components change significantly by nearly three orders of magnitude as the electric field is further increased. A similar enhancement is revealed in our computation of the other QMT components, $g_{xy}$ and $g_{yy}$, as presented in Fig.~\ref{Fig_4layer_QMT} with increasing $E$. Thus, we can tune the magnitudes and the distributions of the QMT components with $E$, which effectively allows tuning the transport properties and the quantum weights. Furthermore, this giant enhancement may allow an easier pathway to directly detect the QMT in future experiments.\\

\begin{figure}[t]
% \centering
\includegraphics[width=0.48\textwidth]{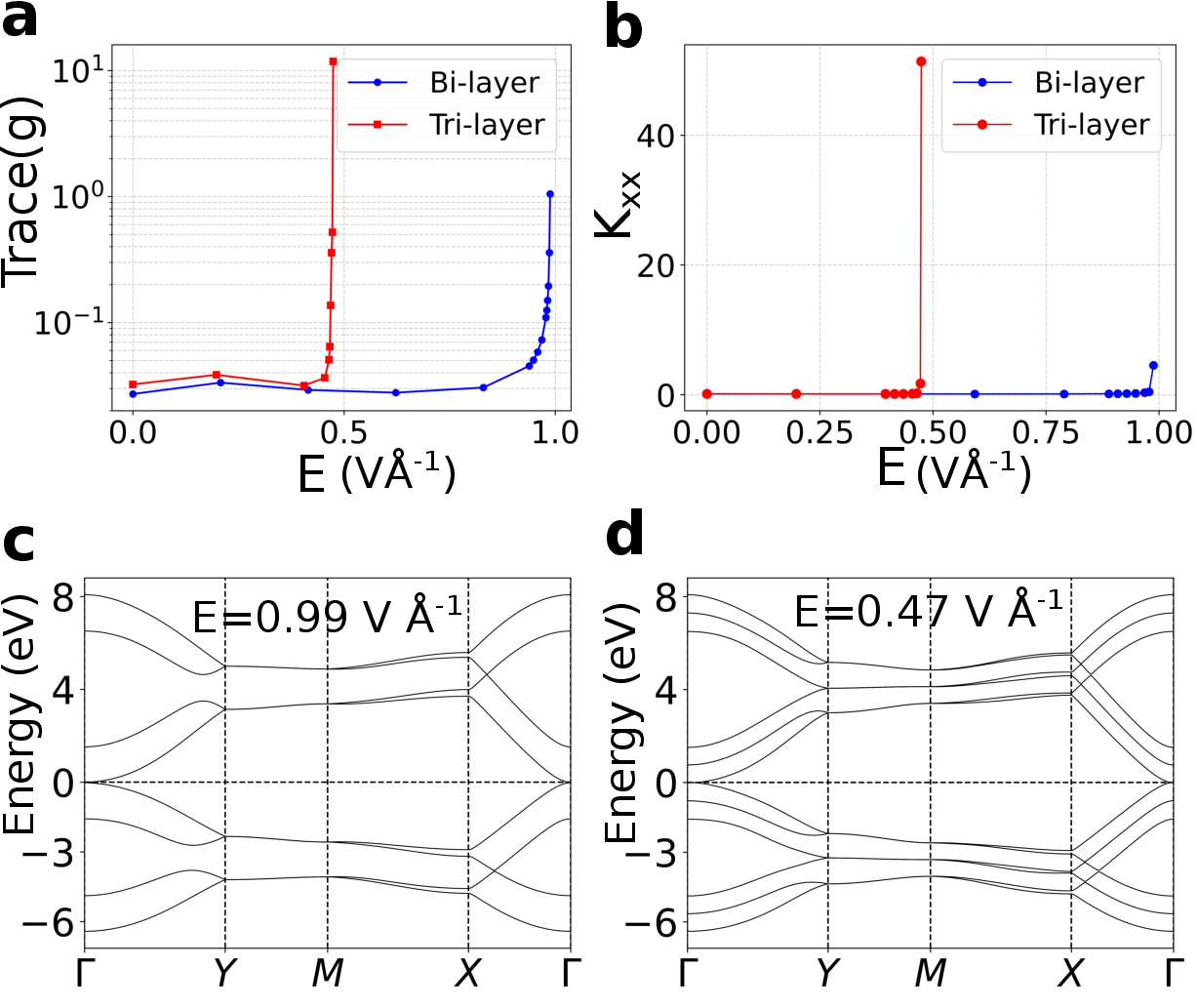}
\caption{\textbf{Quantum metric for bi- and tri-layer phosphorene under varying electric field.} (a) Variation of the trace of the quantum metric with the electric field, $E$, for bi- and tri-layer phosphorene. (b) The quantum weight for bi- and tri-layer phosphorene with the electric field. We note that the magnitude of the trace of the QMT and the quantum weight increases with $E$ and reaches a maximum at the critical fields $E = 0.99~\mathrm{V\,\AA^{-1}}$ and $E = 0.47~\mathrm{V\,\AA^{-1}}$, where the conduction and valence bands touch for the bi-layer and tri-layer cases, respectively. The corresponding critical band structures are shown in panels (c) and (d) for bi- and tri-layer phosphorene, respectively. Thus, the quantum weight and the quantum metric components in can be tuned directly by the applied electric field.}
\label{Fig_Trace_Bi_Tri}
\end{figure}

\noindent\textit{\textbf{Quantum metric in bi-layer and tri-layer phosphorene--}}
Next, we investigate how the layer-dependent band gap tunability under an applied electric field influences the QMT. In particular, we examine the effects of the electric field in bi- and tri-layer phosphorene. In Fig.~\ref{Fig_Trace_Bi_Tri}(a), we present the trace of the quantum metric as a function of the applied field strength. We observe that the trace increases sharply with the electric field $E$ for both bi- and tri-layer cases, after reaching the respective critical fields $E = 0.99~\mathrm{V\,\AA^{-1}}$ and $E = 0.47~\mathrm{V\,\AA^{-1}}$. We also evaluate the quantum weight $K_{xx}$ for the bi- and tri-layer cases, as shown in Fig.~\ref{Fig_Trace_Bi_Tri}(b). It mimics the behavior of the corresponding trace of the QMT, as expected. We find that the other components, $K_{xy}$ and $K_{yy}$, show similar behavior. From the corresponding band structure shown in Fig.~\ref{Fig_Trace_Bi_Tri}(c) and Fig.~\ref{Fig_Trace_Bi_Tri}(d), we note that the band gap closes at these critical field values, leading to a pronounced enhancement of the QMT and quantum weight components. The bi-layer and tri-layer cases exhibit a similar behavior as the quadruple-layer case, with the key difference being that a comparatively higher electric field is required to observe the enhanced quantum metric and quantum weights. Overall, we find that in few-layer phosphorene there exists a layer-dependent tunability of the QMT, in addition to the control provided by the applied electric field.

\noindent \textit{\textbf{Summary and outlook--}}
We have proposed few-layer phosphorene as a promising platform to explore tunable quantum geometry. Our results demonstrate that in this system a band gap tunability can be utilized to control both the magnitude and distribution of the QMT components. The QMT can be tuned either by applying an external electric field, by varying the number of layers, or through a combination of both these effects. Furthermore, we have shown that the quantum weights can also be controlled by modulating the underlying quantum geometry. In summary, our work provides a viable route towards realizing and controlling the quantum metric in experimentally accessible materials. We finally comment on possible experimental exploration of our predictions. Our results can be directly tested using the quasi-quantum metric formalism through circular dichroism angle-resolved photoemission spectroscopy (CD-ARPES), as described in Ref.~\cite{kang2025measurements}. In this technique, the intensity difference between left- and right-circularly polarized light probes the quantum geometric properties of the Bloch states, allowing an experimental reconstruction of the quantum metric in momentum space. It will also be interesting to explore how our system, with broken inversion symmetry in presence of the applied electric field, might connect to the formalism of Kim \textit{et al.} using the notion of pseudospins~\cite{kim2025direct}. Furthermore, our findings pertaining to quantum weights are also experimentally accessible through X-ray scattering or electron energy-loss spectroscopy, since the quantum weight is fundamentally related to the long-wavelength limit of the static structure factor, as shown in Ref.~\cite{onishi2024quantum}. In particular, these techniques probe the density-density correlations in the ground state, which encode the same quantum geometric information captured by the quantum weight. Thus, both CD-ARPES and scattering-based measurements provide complementary experimental routes to validate our theoretical predictions, linking quantum geometry to observable material properties. \\

\noindent \textit{\textbf{Acknowledgments--}}
We thank K. Chatterjee, R. Sarkar, N. B. Joseph, A. Bose, and A. Mondal for useful discussions. M.A.R. is supported by a graduate fellowship of the Indian Institute of Science. A.N. thanks DST CRG grant (CRG/2023/000114) for support.

\section*{APPENDIX}
\renewcommand{\theequation}{A\arabic{equation}}
\renewcommand{\thesection}{A\arabic{section}}
\renewcommand {\thefigure}{A\arabic{figure}}
\setcounter{equation}{0}
\setcounter{figure}{0}

\begin{figure}
\renewcommand{\thefigure}{Appendix\,\arabic{figure}}
\setcounter{figure}{0}
% \centering
\includegraphics[width=0.48\textwidth]{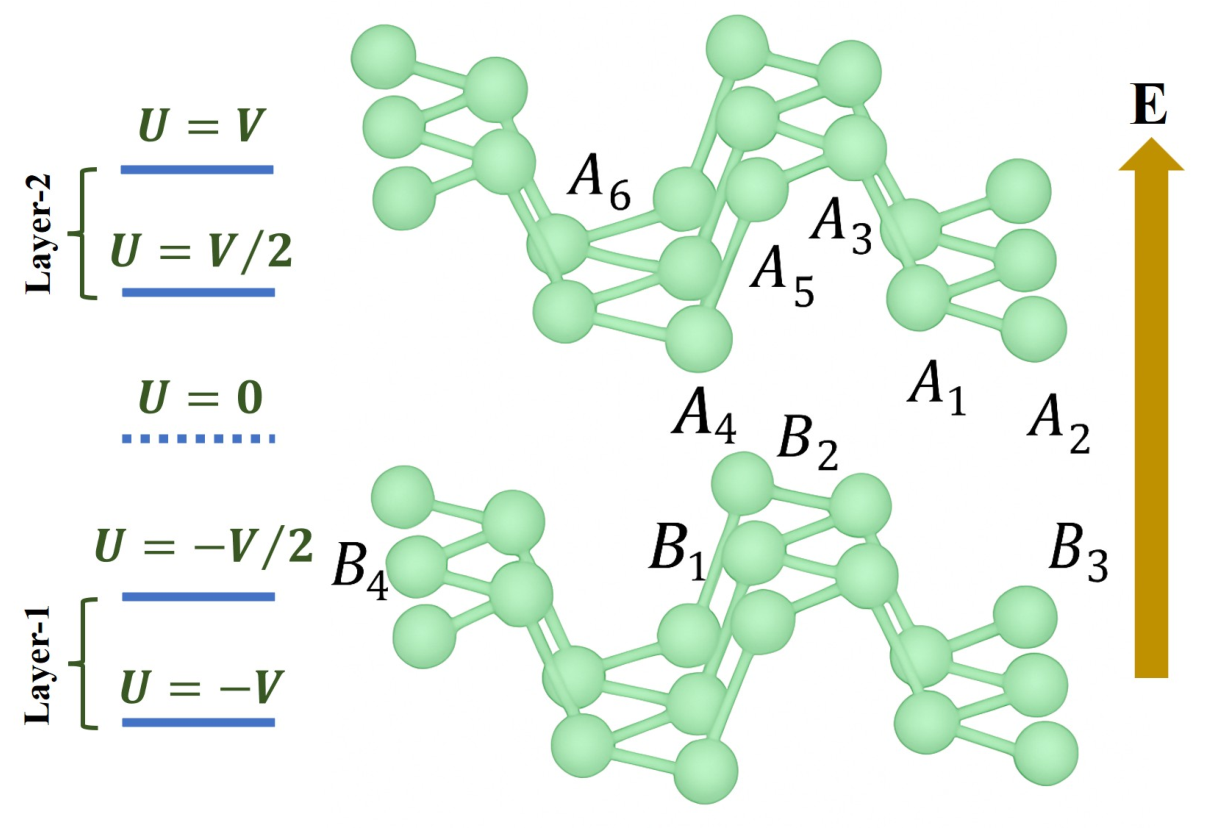}
\caption{\textbf{Hopping model and electrostatic potential distribution under a transverse electric field.} Schematic illustration of bilayer phosphorene showing atomic sites connected by various intralayer and interlayer hopping parameters given in Table-I. The electrostatic potential \( U \) varies between atoms located at different heights along the direction of the applied electric field \( \mathbf{E} \), which is perpendicular to the phosphorene layers. The intralayer hopping amplitudes in phosphorene are $t_{10}^{\parallel}, t_{20}^{\parallel}, t_{30}^{\parallel}, t_{40}^{\parallel},$ and $t_{50}^{\parallel}$, corresponding to the site pairs $A_1-A_2$, $A_1-A_3$, $A_1-A_4$, $A_1-A_5$, and $A_2-A_5$, respectively. For multilayer phosphorene, the interlayer hopping parameters are also present and denoted by $t_{10}^{\perp}, t_{20}^{\perp}, t_{30}^{\perp},$ and $t_{40}^{\perp}$, associated with $A_4-B_1$, $A_6-B_2$, $A_6-B_4$, and $A_4-B_4$, respectively.}
\label{fig:model}
\end{figure}

\subsection{Details of the crystal structure}
The puckered structure of phosphorene gives rise to two distinct types of P--P bonds: in-plane bonds of length $d_1 = 2.22~\text{\AA}$, which form the upper and lower sublayers, and interlayer bonds of length $d_2 = 2.24~\text{\AA}$, which connect these sublayers at an angle of approximately $71.7^\circ$ relative to the plane~\cite{li2014electrons}. The corresponding first Brillouin zone, shown in Fig.~\ref{Fig_4layer_bands_E}(b), highlights the high-symmetry points $\Gamma = (0, 0)$, $X = (\pi/a, 0)$, $M = (\pi/a, \pi/b)$, and $Y = (0, \pi/b)$. Phosphorene preserves the in-plane translational symmetry of bulk black phosphorus, which crystallizes in a base-centered orthorhombic lattice (space group \emph{Cmca}, No.~64). Its nonsymmorphic symmetry is described by the $D_{2h}$ point group, consisting of eight symmetry operations. These include four pure operations -- identity ($E$), inversion ($i$), twofold rotation about $y$ ($C_{2y}$), and reflection through $y=0$ ($R_y$) -- and four operations combining a half-lattice translation $\tau = (a_x/2, a_y/2)$ with rotations or reflections ($\tau C_{2x}$, $\tau C_{2z}$, $\tau R_x$, $\tau R_z$). \\

The application of a transverse electric field induces a potential difference between the layers of phosphorene. In particular, the potential difference between the top of the first layer and the bottom of the second layer (which lies above the first) is denoted by $V$, as shown in Fig.~\ref{fig:model}. It is worth noting that each phosphorene layer is puckered, with one pair of atoms in the unit cell positioned at a slightly different height than the other pair. Since this intra-layer buckling ($\approx$ 2.10 \AA {}) is much smaller than the interlayer separation of 5.3 \AA{}~\cite{shulenburger2015nature}, the resulting potential energy difference between the two atomic sites within a unit cell, $\pm V/2$, is significantly smaller than the potential difference between adjacent layers. \\

\begin{table}[h!]
\centering
\caption{Tight-binding hopping parameters for multilayer phosphorene. Intralayer hopping terms, \( t^{\parallel}_{ij} \), and interlayer hopping strengths, \( t^{\perp}_{ij} \), are listed along with their corresponding atomic pairs and numerical values used in this work. For a visual representation of the atomic sites, please see the Appendix (Fig.~\ref{fig:model}).}
\label{tab:hopping-parameters}
\begin{tabular}{c c c c}
\hline
\hline
Hopping type & Atom pair & Parameter & Value (eV) \\
\hline
\multirow{5}{*}{Intralayer} & $A_1 \leftrightarrow A_2$ & $t^{\parallel}_{10}$ & -1.220 \\
                            & $A_1 \leftrightarrow A_3$ & $t^{\parallel}_{20}$ & 3.665 \\
                            & $A_1 \leftrightarrow A_4$ & $t^{\parallel}_{30}$ & -0.205 \\
                            & $A_1 \leftrightarrow A_5$ & $t^{\parallel}_{40}$ & -0.105 \\
                            & $A_2 \leftrightarrow A_5$ & $t^{\parallel}_{50}$ & -0.055 \\
\hline
\multirow{4}{*}{Interlayer} & $A_4 \leftrightarrow B_1$ & $t^{\perp}_{10}$ & 0.295 \\
                             & $A_6 \leftrightarrow B_2$ & $t^{\perp}_{20}$ & 0.273 \\
                             & $A_6 \leftrightarrow B_4$ & $t^{\perp}_{30}$ & -0.091 \\
                             & $A_4 \leftrightarrow B_4$ & $t^{\perp}_{40}$ & -0.151 \\
\hline
\hline
\end{tabular}
\end{table}

\subsection{Details of the numerical simulation}

The QMT and the Berry curvature can also be expressed in terms of the derivatives of the Hamiltonian as~\cite{kang2025measurements}

\begin{equation}
\begin{aligned}
g^n_{ij}(\mathbf{k}) = \sum_{m \neq n} \mathrm{Re} \left[ 
\frac{ \braket{ \psi_n | \partial_{k_i} H(\mathbf{k}) | \psi_m } 
       \braket{ \psi_m | \partial_{k_j} H(\mathbf{k}) | \psi_n } }
     {(E_m - E_n)^2} 
\right],
\label{Eq_QMT}
\end{aligned}
\end{equation}

\begin{equation}
\begin{aligned}
\Omega^n_{ij}(\mathbf{k}) = -2 \sum_{m \neq n} \mathrm{Im} \left[
\frac{ \braket{ \psi_n | \partial_{k_i} H(\mathbf{k}) | \psi_m } 
       \braket{ \psi_m | \partial_{k_j} H(\mathbf{k}) | \psi_n } }
     {(E_m - E_n)^2} 
\right].
\label{Eq_BC}
\end{aligned}
\end{equation}

where, $i$, $j$ represent the spatial indices, \( \partial_{k_i} = \frac{\partial}{\partial k_i} \), $n$ is the band index, and $E_n$ is the eigenvalue of $n$-th band. 
We first construct a realistic tight-binding model in the momentum space for few-layer phosphorene under external field using PythTB~\cite{coh2022pythtb}. The corresponding Hamiltonian obtained from this model serves as the input for our numerical calculations of the quantum metric components $g_{xx}$, $g_{xy}$, and $g_{yy}$ for each band, following Eq.~\ref{Eq_QMT}. The Hamiltonian-derivative approach offers certain numerical advantages over the direct evaluation of Eq.~\ref{Eq_QGT}, as it avoids the explicit computation of derivatives of the Bloch wavefunctions. These quantities are evaluated on a mesh over the Brillouin zone for various values of the applied electric field.

\bibliography{ref.bib}
\end{document}